\documentclass{llncs}
\usepackage{setspace}
\usepackage{graphics}
\usepackage{graphicx}
\usepackage{amssymb}
\newcommand{\mo}{{\it mathoverflow }}
\newcommand{\minip}{{\it mini-polymath }}
\author{Ursula Martin and Alison Pease}
\title{Mathematical practice, crowdsourcing, and social machines}
\date{Draft: \today}
\institute{
Queen Mary University of London\\
\email{Ursula.Martin@qmul.ac.uk, Alison.Pease@eecs.qmul.ac.uk}}

\begin{document}
\maketitle

\begin{abstract} 

The highest level of mathematics has traditionally been seen as a solitary endeavour, to produce a proof for review and acceptance by research peers. Mathematics is now at a remarkable inflexion point, with new technology radically extending the power and limits of individuals. Crowdsourcing pulls together diverse experts to solve problems; symbolic computation tackles huge routine calculations; and computers check proofs too long and complicated for humans to comprehend. 

The {\em Study of Mathematical Practice} is an emerging interdisciplinary field which draws on philosophy and social science  to understand how mathematics is produced.  Online mathematical activity  provides a novel and rich source of data for empirical investigation of mathematical practice - for example the community question-answering system {\it mathoverflow} contains around 40,000 mathematical conversations, and {\it polymath} collaborations provide transcripts of the process of discovering proofs. Our preliminary  investigations have demonstrated the importance of ``soft'' aspects such as analogy and creativity, alongside deduction and proof, in the production of mathematics, and have given us new ways to think about the  roles of people and machines in creating new mathematical knowledge. We discuss  further investigation of these resources and what it might reveal.

Crowdsourced mathematical activity is an example of a ``social machine'', a new paradigm, identified by Berners-Lee, for viewing a combination of people and computers as a single problem-solving entity, and the subject of major international research endeavours.  We outline a future research agenda for mathematics social machines, a combination of people, computers, and mathematical archives to create and apply mathematics, with the potential to change the way people do mathematics, and to transform the reach, pace, and impact of mathematics research.
\end{abstract}
\section{Introduction}
For centuries, the highest level of mathematical research has been seen as an isolated creative activity, whose goal is to identify mathematical truths, and  justify them by rigorous logical arguments which are presented for review and acceptance by research peers. 

Yet  mathematical discovery also involves ÒsoftÓ aspects such as creativity, informal argument, error and analogy.  For example, in an interview in 2000 \cite{Wilesinterview} Andrew Wiles describes his 1989 proof of Fermat's theorem in almost mystical terms ``... and sometimes I realized that nothing that had ever been done before was any use at all. Then I just had to find something completely new; it's a mystery where that comes from."  Michael Atiyah remarked at a workshop in Edinburgh in 2012 \cite{Edi2012}``I make mistakes all the time" and ``I published a theorem in topology. I didn't know why the proof worked, I didn't understand why the theorem was true. This worried me. Years later we generalised it---we looked at not just finite groups, but Lie groups. By the time we'd built up a framework, the theorem was obvious. The original theorem was a special case of this. We got a beautiful theorem and proof.''

Computer assisted proof formed some of the earliest experiments in artificial intelligence: in 1955 Newell, Shaw and Simon's Logic Theorist searched forward from axioms to look for proofs of results taken from Russell and Whitehead's 1911 Principia Mathematica. Simon reported in a 1994 interview \cite{Simon} that he had written to  Russell  (who died in 1970, aged 97), who ``wrote back that if we'd told him this earlier, he and Whitehead could have saved ten years of their lives. He seemed amused and, I think, pleased."  By the mid-1980s a variety of approaches and software tools, such as the theorem provers HOL, NuPrl and Nqthm, had started to be developed for practical reasoning about programs: \cite{DBLP:journals/annals/Jones03} is a thorough account of the early history.  This laid the foundation for  a flourishing academic and industry community, and currently verification to ensure error-free systems is a major endeavour in companies like Intel and Microsoft  \cite{DBLP:books/daglib/0022394}, as well as supporting specialist small companies.   At the same time theorem provers are now being used by an influential community of mathematicians. Tom Hales and his team have almost completed a ten-year formalisation of their proof of the Kepler conjecture, using several theorem provers to confirm his major 1998 paper \cite{DBLP:journals/dcg/HalesHMNOZ10}. In September 2012 Georges Gonthier announced that after a six year effort his team had completed a formalisation, in the Coq theorem prover, of one of the most important and longest proofs of 20th century algebra, the 255 page odd-order theorem \cite{DBLP:conf/popl/Gonthier13}. He summarised the endeavour as: 
\begin{quote}
Number of lines \~{} 170 000\\
Number of definitions \~{} 15 000\\
Number of theorems \~{} 4 200\\
Fun \~{} enormous!
\end{quote}
The growth in the use of computers in mathematics, and in particular of computer proof, has provoked debate, reflecting the contrast between the ``logical'' and ``human'' aspects  of creating mathematics: see  \cite{MR1698135} for a survey. For example in an influential paper in 1979, De Millo, Lipton and Perlis  \cite{DBLP:journals/cacm/DeMilloLP79}, argued that ``Mathematical proofs increase our confidence in the truth of mathematical statements only after they have been subjected to the social mechanisms of the mathematical community'', and expressed concern over ``symbol chauvinism".  Similar concerns were raised in the mathematical community over the use of a computer by Appel and Haken \cite{MR0543796} to settle the long standing four colour conjecture.
Indeed, Hume,  in his 1739 Treatise on Human Nature \cite{hume2003treatise} p231, identified the importance of the social context of proof:
\begin{quote}
There is no Algebraist nor Mathematician so expert in his science, as to place entire confidence in any truth immediately upon his discovery of it, or regard it as any thing, but a mere probability. Every time he runs over his proofs, his confidence encreases; but still more by the approbation of his friends; and is rais'd to its utmost perfection by the universal assent and applauses of the learned world. [{\em sic}]
\end{quote}

The sociology of science addresses such paradoxes in the understanding of the scientific process, and a comprehensive account  is given by sociologist Donald MacKenzie in his 2001 book ``Mechanizing Proof'' \cite{MR2039787}. He concludes that used to extend human capacity the computer is benign, but that ``trust in the computer cannot entirely replace trust in the human collectivity''.  In recent years ``the study of mathematical practice''  has emerged from the work of P\'olya and Lakatos as a subdiscipline drawing upon the  work of sociologists, cognitive scientists, philosophers and the narratives of mathematicians themselves, to study exactly what it is that mathematicians do to create mathematics. Section 2 of this paper contains a fuller account.

The mathematical community were ``early adopters'' of the internet for disseminating papers, sharing data, and blogging, and in recent years have developed systems for ``crowdsourcing'' (albeit among a highly specialised crowd) the production of mathematics through  collaboration and sharing, providing further evidence for the social nature of mathematics.  To give just a few examples:
\begin{itemize}
\item A number of senior mathematicians produce influential and widely read blogs.  In the summer of 2010 a paper was released plausibly claiming to prove one of the major challenges of theoretical computer science, that $P \not=NP$: it was withdrawn after rapid analysis organised by senior scientist-bloggers, and coordinated from Richard Lipton's blog. Fields Medallist Sir Tim Gowers used his blog to lead an international debate  about mathematics publishing.  
\item {\it polymath} collaborative proofs, a new idea led by  Gowers, use a blog and wiki for collaboration among mathematicians from different backgrounds  and have led  to major advances \cite{gowers2009massively}
\item  discussion fora allow rapid  informal interaction and problem-solving;  in three years the  community question answering system for research mathematicians {\it mathoverflow} has 23,000 users and has hosted 40,000  conversations
\item  the widely used ``Online Encyclopaedia of Integer Sequences" (OEIS) invokes subtle pattern matching against over 200,000 user-provided sequences on a few digits of input to propose matching sequences: so for example input of (3 1 4 1) returns $\pi$ (and other possibilities) \cite{OEIS}  
\item the {\it arXiv} holds around 750K preprints in mathematics and related fields. By providing open access ahead of journal submission, it has markedly increased the speed of refereeing, widely identified as a bottleneck to the pace of research \cite{MR2856147}
\item Innocentive \cite{innocentive}, a site hosting open innovation and crowdsourcing challenges, has hosted around 1,500 challenges with a 57\% success rate, of which around 10\% were tagged as mathematics or ICT.
\end{itemize}

As well as having a remarkable effect on mathematical productivity, these systems provide substantial and unprecedented evidence  for studying mathematical practice,  allowing the augmentation of   traditional ethnography with a variety of empirical techniques for analysing the texts and network structures of the interactions. In Section 3 we describe  two of our own recent preliminary studies, of \mo and {\it polymath}, which provide evidence for the theories of P\'olya and Lakatos, and shed new light on mathematical practice, and on the  current or future computational tools that might enhance it. Analysing the content of a sample of questions and responses, we find that {\it mathoverflow} is very effective, with 90\% of our sample of questions answered completely or in part. A typical response is an informal dialogue, allowing error and speculation, rather than rigorous mathematical argument: a surprising 37\% of our sample discussions acknowledged error. Looking at one of the recent \minip problems, we find  only 24\% of the 174 comments formed the development of the final proof, with the remainder comprising a high proportion of examples (33\%) alongside conjectures and social glue. 
We conclude that extending the power and reach of {\it mathoverflow} or {\it polymath} through a combination of people and machines raises new challenges  for artificial intelligence and computational mathematics, in particular how to handle error, analogy and informal reasoning.

Of course, mathematics is not the only science in which productive  new human collaborations are made possible by machines. Over the past twenty years researchers in e-science have devised systems such as Goble's myExperiment~\cite{DBLP:journals/software/RoureG09}  for managing scientific workflow, especially  in bioinformatics, so that data, annotations, experiments, and results can be documented and shared across a uniform platform, rather than in a mixture of stand alone software systems and formats. Michael Nielsen, one of the founders of {\it polymath}, in his 2011 book ``Reinventing discovery'' \cite{nielsen2011reinventing} discusses a number of examples of crowdsourced and citizen science. Alongside {\it polymath}, he describes Galaxy Zoo, which allows members of the public to look for features of interest in images of galaxies, and has led to new discoveries, and Foldit,  an online game where users solve protein folding problems. 

Considered more broadly, such systems are exemplars of ``Social machines'', a broad new paradigm identified by Berners-Lee in his 1999 book ``Weaving the Web'' \cite{DBLP:books/daglib/0020403}, for  viewing a combination of people and computers as a single problem-solving entity. Berners-Lee describes a dream of collaborating through shared knowledge: \begin{quote}
Real life is and must be full of all kinds of social constraint --- the very processes from which society arises. Computers can help if we use them to create abstract social machines on the Web: processes in which the people do the creative work and the machine does the administration. . . The stage is set for an evolutionary growth of new social engines. The ability to create new forms of social process would be given to the world at large, and development would be rapid.
\end{quote}
Current social machines  provide platforms for sharing knowledge and leading to innovation, discovery, commercial opportunity or social benefit:  the combination of mobile phones, Twitter and google maps used to create real-time maps of the effects of natural disasters has been a motivating example. Future more ambitious social machines will combine social involvement and sophisticated automation, and are now the subject of major research, for example in Southampton's SOCIAM project \cite{SOCIAM} following an agenda laid out by Hendler and Berners-Lee \cite{DBLP:journals/ai/HendlerB10}.    In Section 4 we look at collaborative mathematics systems through the lens of social machines research, presenting a research agenda that further develops the results of work on the practice of mathematics.

\section{The study of mathematical practice}
The study of mathematical practice emerged as a fledgling discipline in the 1940's when
mathematician and educator Georg P\'olya formulated problem-solving heuristics
designed to aid mathematics students. These heuristics, such as
``rephrase the question'', and ``draw a diagram'' were based on P\'olya's intuition
about rules of thumb which he himself followed during his
research, and have been influential in mathematics education
(although not without critics, who argue that meta-heuristics are
needed to determine when a particular route is likely to be fruitful
\cite{larsen,nunokawa,schoenfeld}). P\'olya's idea, that it is possible to identify heuristics which
describe mathematical research -- a logic of discovery -- was extended
by Imre Lakatos, fellow countryman and philosopher of mathematics and
science.\footnote{Lakatos translated P\'olya's \cite{pol} and other mathematical works into Hungarian before
developing his own logic of discovery, intended to carry on where
P\'olya left off \cite[p. 7]{lakatos}.} Lakatos used in-depth analyses of extended
historical case studies to formulate patterns of reasoning which
characterised conversations about a mathematical conjecture and its
proof. These patterns focused on interactions between mathematicians
and, in particular, on the role that counterexamples play in driving
negotiation and development of concepts, conjectures and proofs. 

\begin{figure}
\begin{center}
\includegraphics[width=.5\linewidth]{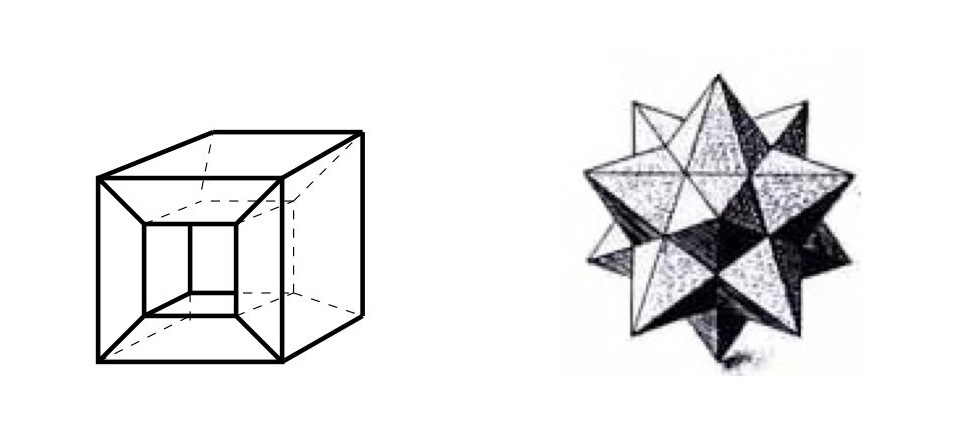}
\caption{Controversial polyhedra: A picture-frame, on the left, for which V - E + F
  = 16 - 32 + 16 = 0, and Kepler's star-polyhedron, on the right, for
  which V - E + F can be 12 - 30 + 12 = -6 (if it has star-pentagon faces) or  32 - 90 +60 = 2 (if it has triangular faces).}
\label{polys}
\end{center}
\end{figure}

Lakatos demonstrated his argument by presenting a rational
reconstruction of the development of Euler's conjecture that for any
polyhedron, the number of vertices (V) minus the number of edges (E)
plus the number of faces (F) is equal to two; and Cauchy's proof of
the conjecture that the limit of any convergent series of continuous
functions is itself continuous. He outlined six methods for modifying
mathematical ideas and guiding communication: surrender,
monster-barring, exception-barring, monster-adjusting,
lemma-incorporation, and proofs and refutations. These are largely
triggered by counterexamples, or problematic entities, and result in a
modified proof, conjecture or concept. For instance, the methods of
{\em monster-barring} and {\em monster-adjusting} exploit ambiguity or vagueness in concept
definitions in order to attack or defend a conjecture, by (re)defining
a concept in such a way that a problematic object is either excluded
or included. With {\em monster-barring}, the ambiguous concept is central to the
conjecture and defines the domain of application, such as a
``polyhedron'' (in Euler's conjecture), a ``finite group'' (in Lagrange's theorem), or an ``even
number'' (in Goldbach's conjecture). Here, Lakatos presents
the picture-frame, for which V - E + F = 16 - 32 + 16 =
0 (see figure
\ref{polys}): this is ``monster-barred'' as being an invalid example of a
polyhedron, and the definition of polyhedron tightened to exclude
it. With {\em monster-adjusting}, the ambiguous concept is a
sub-concept (appears in the {\em definition} of the central concept),
such as ``face'', ``identity'', or ``division'' (following the polyhedron/finite group/even
number examples). (Re)defining this sub-concept can provide an
alternative way of viewing a problematic object in such a way that it
ceases to be problematic:   Lakatos gives the example of Kepler's
star-polyhedron, which is a counterexample if V - E + F is 12 - 30 +
12 = -6 (where its faces are seen as
star-pentagons), but can be salvaged if we see  V - E + F as 32 - 90
+60 = 2 (where its faces are seen as triangles) (see figure
\ref{polys}). The result of both of these
methods is a preserved conjecture statement, where the meaning of
the terms in it have been revised or clarified.

In Lakatos's exception-barring method, a counterexample is seen as an
exception, triggering a refinement to the conjecture, and in his
lemma-incorporation and proofs and refutations methods, problematic
objects are found and examined to see whether they are
counterexamples to a conjecture or a proof step, which are then
revised accordingly. 

Lakatos held an essentially optimistic view of mathematics, in which
the process of mathematics traditionally thought of as impenetrable
and inexplicable by rational laws, considered to be lucky guess work
or intuition, is seen in a rationalist light, thereby opening up new
arenas of rational thought.  He challenged Popper's view
\cite{popper} that philosophers can form theories about how to
evaluate conjectures, but not how to generate them, which should be
left to psychologists and sociologists. Rather, Lakatos believed that
philosophers could theorise about both of these aspects of the
scientific and mathematical process.  He challenged Popper's view in
two ways - arguing that {\em (i)} there {\em is} a logic of discovery,
the process of generating conjectures and proof ideas {\em is} subject
to rational laws; and {\em (ii)} the distinction between discovery and
justification is misleading as each affects the other; {\em i.e.}, the way in
which we discover a conjecture affects our proof (justification) of
it, and proof ideas affect what it is that we are trying to prove (see
\cite{larvor}).  This happens to such an extent that the boundaries
of each are blurred. These ideas have a direct translation into
automated proof research, suggesting that conjecture and concept
generation are subject to rationality as well as proof, and therefore systems
can (perhaps even should) be developed which integrate these
theory-development aspects alongside proof generation. 

At the heart of both
P\'olya and Lakatos's work was the idea that the mechanisms by which
research mathematics progresses -- as messy, fallible, and speculative as this
may be -- can usefully be studied via analysis of informal
mathematics. This idea has been welcomed and extended by a variety of
disciplines; principally philosophy, history sociology, cognitive science and
mathematics education \cite{aberdein,cellucci,corfield,mancosu}.
The development of computer support for mathematical reasoning provides further motivation for studying the
processes behind informal mathematics, particularly in the light of the criticisms this has sometimes received.  Sociologist Goffman
\cite{goffman} provides a useful distinction here, of front and backstage activities,
where activities in the front are services designed for public
consumption, and those in the back constitute the private preparation
of the services. Hersh \cite{hersh:91} extends this distinction to mathematics, where
textbook or publication-style ``finished mathematics'' takes frontstage, and the
informal workings and conversations about ``mathematics in the making''
is hidden away backstage. P\'olya employed a similar distinction, and Lakatos warned of the dangers of hiding the backstage process, either from students (rendering the subject
 impenetrable) or from experts (making it more difficult to develop concepts
 or conjectures which may arise out of earlier versions of a theorem
 statement). Computer support for mathematics, such as computer algebra or computational mathematics, has  typically been for the frontstage. A second, far less developed, approach is to focus on the
backstage, including the mistakes, the dead ends and the unfinished,
and to try to extract principles which are sufficiently clear as to
allow an algorithmic interpretation: the study of
mathematical practice  provides a starting point for this work.



Implicit or explicit in much work on mathematical practice is the recognition that
mathematics takes place in a social context. Education theorist, Paul
Ernest \cite{ernest:97}, sees mathematics as being socially constructed via
conversation; a conversation which is as bound by linguistic and
social conventions as any other discourse. Thus, if such conventions are
violated (by other cultures, or, perhaps, by machines) then shared
understanding is lost and --  mirroring Kuhnian paradigm
shift -- new conventions may need to be formed which
accommodate the rogue participant. Kitcher \cite{kitcher:83}, a philosopher of mathematics, elaborates what a mathematical practice might
mean, suggesting a socio-cultural definition as consisting in a
language and four socially negotiated sets: accepted statements, accepted reasonings,
questions which are considered to be important and meta-mathematical
views such as standards of proof and the role of mathematics in
science (agreement over the content of these sets helps to define a
mathematical culture). Mackenzie  \cite{MR2039787} looked at the role of proof, especially computer proof, and   his student Barany \cite{barany:chalk} used ethnographic methods to trace the cycle of development and flow of
mathematical ideas from informal thoughts, to seminar, to publication,
to dissemination and classroom, and back to informal
thoughts. He sees (re)representations in varying media such as notes,
blackboard scribbles, physical manifestations or patterns of items on a
desk, as necessary, for the knowledge to be decoded and encoded into
socially and cognitively acceptable forms. In particular, Barany
investigated the relationship between the material (the ``pointings,
tappings, rubbings, and writings'' of mathematics \cite[p.9]{barany:chalk}) and the abstract, arguing
that each constrains the other. Other developments in the study of mathematical practice include work
on visualisation, such as diagrammatic reasoning in mathematics
\cite{giaquinto:07,mancosu:05}; analogies, such as between
mathematical theories and axiom sets
\cite{bartha:10,Sch08}; and mathematical concept development, such as
ways to determine potential fruitfulness of rival definitions
\cite{tappenden08a,tappenden08b}.  At the heart of many of these analyses lies
the question of what proof is for, and the recognition that it plays
multiple roles; explaining, convincing, evaluating, aiding memory, and
so on, complementing or replacing traditional notions of proof as a
guarantee of truth). This in turn gives an alternative  picture of  machines as members of a mathematical community.

\section{Mathematical practice and crowdsourced mathematics}

In this section we outline preliminary results from our own ongoing programme of work which uses collaborative online systems  as an evidence base for further understanding of mathematical practice. We studied  a sample of  \mo questions and the ensuing discussions \cite{MartinSOHUMAN}, and the third \minip problem \cite{MarPea12b},  looking at the kinds of activities taking place,  the relative importance of each, and evidence for  theories of mathematical practice described in the previous section, especially the work of  P\'olya \cite{pol} and Lakatos \cite{lakatos}.

\mo and {\it polymath} are  similar in that they are examples of the backstage of collaborative mathematics. They provide records of mathematicians collaborating through nothing more than conversation, underpinned by varying levels of shared expertise and context. While participants may invoke results from computational engines, such as GAP or Maple, or cite the literature, neither system contains any formal links to software or databases.   The usual presentation of mathematics in research papers is the frontstage, in a standardised precise and rigorous style:  for example,  the response to a conjecture is either a counterexample, or a proof of a corresponding theorem, structured by means of intermediate definitions, theorems and proofs.  By contrast  these systems present the backstage of mathematics: facts or short chains of inference  that are relevant to the question, but may not answer it directly,  justified by reference to mathematical knowledge that the responder expects the other participants to have.

\subsection{Mathoverflow}
\begin{figure}
\begin{center}
\includegraphics[width=.9\linewidth]{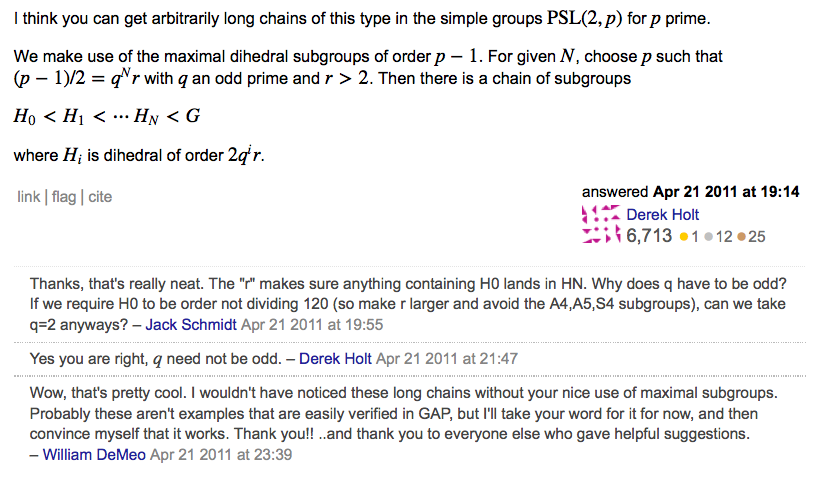}
\caption{A typical \mo conversation}
\label{moq}
\end{center}
\end{figure}

Discussion fora for research mathematics have evolved from the early newsnet newsgroups to modern systems based on the {\it stackexchange} architecture, which allow rapid  informal interaction and problem-solving. In three years {\it mathoverflow.net} has accumulated 23,000 users and hosted 40,000  conversations. Figure \ref{moq} shows part of a \mo conversation \cite{moexample}, in answer to a question about the existence of certain kinds of chains of subgroups. The highly technical nature of research mathematics means that,  in contrast to activities like GalaxyZoo, this is not currently an endeavour accessible to the public at large: a separate site  {\it math.stackexchange.com} is a broader question and answer site ``for people studying math at any level and professionals in related fields''.  Within \mo, house rules give detailed guidance, and stress clarity, precision, and asking questions with a clear answer. Moderation is fairly tight, and  some complain it constrains discussion.  

The design of such systems has been subject to considerable analysis (see, for instance, \cite{begel}), and {\it meta.mathoverflow} contains many reflective discussions.  A key element is user ratings of questions and responses, which combine to form reputation ratings for users. These have been studied by psychologists Tausczik and Pennebaker \cite{DBLP:conf/chi/TausczikP11,DBLP:conf/cscw/TausczikP12}, who  concluded that \mo reputations offline (assessed by numbers of papers published) and in \mo  were consistently and independently related to the \mo ratings of authors' submissions, and that while more experienced contributors were more likely to be motivated by a desire to help others, all were motivated by building their \mo  reputation. 

%

Within   \mo we identified the predominant kinds of questions as: {\bf Conjecture (36\%)}, which ask whether or under what circumstances a statement    is true; {\bf What is this (28\%)}, which  describe an object or phenomenon and ask what is known about it; and 
{\bf Example (14\%)} which ask for examples of a phenomenon or an object with particular properties. Other smaller categories  
ask for an explicit formula or computation technique, for  alternatives to a known proof , for literature references, for help in understanding a difficulty or apparent contradiction,\footnote{Several  questions concerned why Wikipedia and a published paper  seemed to contradict each other.} or for motivation. 

Analysing the answers in our sample shed further light on how the system was being used.   \mo is very effective,  with 90\% of our sample successful, in that they received responses that the questioner flagged as an ``answer'', of which 78\% were reasonable answers to the original question, and a further 12\% were partial or helpful responses that moved knowledge forward in some way.   The high success rate suggests  that, of the infinity of possible mathematical questions,  questioners are becoming adept at choosing those for \mo  that are amenable to its approach.  

The presentation is often speculative and  informal, a style which would have no place in a research paper, reinforced by conversational devices that are accepting of error and invite challenge, such as ``I may be wrong but...'', ``This isn't quite right, but roughly speaking...''.  Where errors are spotted, either by the person who made them or by others, the style is to politely accept and correct them: corrected errors of this kind were found in 37\% of our sample.\footnote{This excludes ``conjecture'' questions where the responses  refutes the conjecture. We looked at {\it discussions} of error: we have no idea how many actual errors there are!}

In 34\% of the responses explicit examples were given, as evidence for, or counterexamples to, conjectures: thus playing exactly the role envisaged by Lakatos.   We return to this below. In 56\% of the responses we found citations to the literature. This includes both finding papers that questioners were unaware of, and extracting results that are not explicit in the paper, but are  straightforward (at least to experts) consequences of the material it contains. 


 It is perhaps worth commenting on things that we did not see. As we shall see in the next section, in developing ``new'' mathematics considerable effort is put into the formation of new concepts and definitions: we saw little of this in {\it mathoverflow}, where questions by and large concern extending or refining existing knowledge and theories.   We see little serious disagreement in our \mo sample: perhaps partly because of the effect of the ``house rules'', but also because of the style of discussion, which is based on evidence from the shared research  background and knowledge of the participants: there is more discussion and debate in {\it meta.mathoverflow}, which  has a broader range of non-technical questions about the development of the discipline and so on. 
 
\subsection{Polymath}
In 2009 the mathematician Timothy Gowers asked ``Is
massively collaborative mathematics possible?''  \cite{is-massively-collaborative-mathematics-possible}, and with 
Terence Tao initiated experiments which invited contributions on a blog to solving open, difficult conjectures.  Participants were asked to 
 follow guidelines \cite{polymath-rules}, which had emerged from an online collaborative discussion, and were intended to encourage widespread participation and
a high degree of interaction, with results arising from the
rapid exchange of informal ideas, rather than parallelisation of
sub-tasks. These included ``It's OK for a mathematical
thought to be tentative, incomplete, or even incorrect'' and ``An
ideal polymath research comment should represent a `quantum of
progress' ''.  While \mo is about asking questions, where typically the questioner believes others in the community may have the answer, {\it polymath} is about collaborating to solve open conjectures.

There have now been seven Polymath discussions, with some still ongoing,  leading to significant advances and published papers, under the byline of ``D J H Polymath'' \cite{MR2912706}. Analysis by Gowers \cite{gowers2009massively}, and by HCI researchers Cranshaw and Kittur \cite{cranshaw},  indicates that 
 {\it polymath} has enabled a level of collaboration which, before the internet,
would probably have been impossible to achieve; the open invitation
has widened the mathematical community; and the focus on
short informal comments has resulted in a readily available and public
record of mathematical progress. As noted by Gowers, this provided ``for possibly the first time ever (though I may well be wrong about this) the
first fully documented account of how a serious research problem was
solved, complete with false starts, dead ends etc.'' \cite{polymath1-and-open-collaborative-mathematics}. 
Four annual \minip projects (so far) have selected   problems from the current International Mathematical Olympiad:  thus in contrast to the open-ended research context of {\it polymath}, participants  trust the question to be solvable without advanced mathematical knowledge. 

We investigated  \minip 3, which used the following problem.
\begin{center}
\begin{footnotesize}
\parbox{14cm}
{
\line(2,0){400}\\
Let $S$ be a finite set of at least two points in the
plane. Assume that no three points of $S$ are collinear. A {\em windmill} is a
process that starts with a line $l$ going through a single point $P
\in S$. The line rotates clockwise about the pivot $P$ until the first time that the
line meets some other point $Q$ belonging to $S$. This point $Q$ takes over
as the new pivot, and the line now rotates clockwise about $Q$, until it
next meets a point of $S$. This process continues indefinitely.

Show that we can choose a point $P$ in $S$ and a line $l$ going
through $P$ such that the resulting windmill uses each point of $S$ as
a pivot infinitely many times.\\ 
\line(2,0){400}
}

\end{footnotesize}
\end{center}
It was solved over a period of 74 minutes by 27 participants through 174 comments on 27 comment threads. People mostly followed the rules, which were largely self regulating due to the speed of responses: a  long answer in response to an older thread was likely to be ignored as the main discussion had moved on.  Some sample comments included: 
\begin{center}
\begin{footnotesize}
\parbox{14cm}{
\line(2,0){400}\\
{\bf 1} If the points form a convex polygon, it is easy.  \\

{\bf 2} Can someone give me *any* other example where the windmill
cycles without visiting all the points? The only one I can come up
with is: loop over the convex hull of S\\

{\bf 3} One can start with any point (since every point of S should be pivot infinitely often), the direction of line that one starts with however matters!  \\

{\bf 4} Perhaps even the line does not matter! Is it possible to prove that any point and  line will do?  \\

{\bf 5} The first point and line $P_0$, $l_0$ cannot be chosen so that $P_0$ is on the boundary of the convex hull of S and $l_0$ picks out an adjacent point on the convex hull. Maybe the strategy should be to take out the convex hull of S from consideration; follow it up by induction on removing successive convex hulls.  \\

{\bf 6} Since the points are in general position, you could define ``the wheel of p'', w(p) to be radial sequence of all the other points p!=p around p. Then, every transition from a point p to q will ``set the windmill in a particular spot'' in q. This device tries to clarify that the new point in a windmill sequence depends (only) on the two previous points of the sequence. \\

\line(2,0){400}}
\end{footnotesize}
\end{center}

 Within \minip 3, we classified the main activity of each of the 174 comments as either:
\begin{center}
\begin{footnotesize}
\parbox{14cm}{
\line(2,0){400}\\
{\bf Example 33\%}  (1, 2 above). Examples and counterexamples played a key role: in understanding and exploring the problem, in clarifying explanations, and in exploring concepts and conjectures about the problem. In  the early stages of understanding the problem, a number of participants were misled by the use of the term  ``windmill'' to think of the rotating line as a half-line, a misunderstanding that led to counterexamples to the result they were asked to prove.\footnote{A line extends indefinitely in both directions, whereas a half-line, or ray, starts at a fixed point and extends indefinitely in one direction only.} \\

{\bf Conjecture 20\%}  (3, 4 above). This category included exploration of the limits of the initial question and various sub-conjectures. We identified conjectures made by analogy; conjectures that generalised the original problem; sub-conjectures towards a proof; and conjectured properties of the main windmill concept.
\\

{\bf Proof 14\%}  (5 above) Proof strategies found included induction, generalisation,  and analogy. 
\\

{\bf Concept 10\%}  (6 above) As well as standard concepts from Euclidean geometry and the like, even in such a relatively simple proof, new concepts arise by analogy; in formulating conjectures; or from considering examples and counterexamples. For example, analogies involving ``windmills'' led to the misapprehension referred to above. 
\\

{\bf Other 23\% } These typically concerned cross referencing to other comments; clarification; and social interjections, both mathematically
interesting and purely social, including smiley faces and the like. All help to create a 
friendly, collaborative, informal and polite environment.\\
\line(2,0){400}}
\end{footnotesize}
\end{center}

\subsection{What do we learn about mathematical practice?}

Both \mo and \minip provide living examples of the backstage of mathematics.

While the utility of P\'olya's  ideas  in an educational setting has been contested,  \minip shows many examples of his problem-solving heuristics operating in a collaborative, as opposed to individual, setting: for example we see participants rephrasing the question, using case splits and trying to generalise the problem. This is hardly surprising, as the questions themselves may have been designed to be solved by these techniques. 

Both \mo and \minip afford precisely the sort of openness that Lakatos advocated in the teaching and presentation of mathematics (described above). We have seen the striking number of examples used in both \mo and \minip: this accords with the emphasis which Lakatos placed on examples. He emphasised fallibility and ambiguity in mathematical development, addressing semantic change
in mathematics as the subject develops, the role that counterexamples play in concept, conjecture and proof development, and the social component of mathematics via a dialectic of ideas. Although his theory was highly social, it was not necessarily collaborative. For reasons of space we single out here Lakatos's notion of ``monster-adjusting'' examples: others are considered in \cite{MarPea12b}.


Monster-adjusting occurs when an object is seen as a supporting example of a conjecture by one person and as a counterexample by another; thus exposing two rival interpretations of a concept definition.  The object then becomes a trigger for concept development and clarification. Thus in our \mo example this occurs, relative to the larger conversation not displayed,  in the  comment and adjustment of Figure \ref{moq} around ``Why does $q$ have to be odd?'' In our \minip study the monster-adjusting occurs in clarifying the rotating line of the question as a full line not a half-line:  the problematic object is an equilateral triangle with 
one point in the centre; this exposes different interpretations of the concept of the rotating line.  

While with sufficient ingenuity most of the examples we found in both systems could be assigned to one or more of Lakatos's categories, the process is quite subtle, and  dependent on context in a way not always taken into account in Laktos's work: the  \mo example taken alone could also be  seen a variation of Lakatos's exception-barring, where the conjecture is strengthened by lifting unnecessary conditions.

While Lakatos identifies the role that hidden assumptions
play, and suggests ways of diagnosing and repairing flawed
assumptions, he does not suggest how they might arise. Here we can go beyond Lakatos and hypothesise as
to what might be the underlying reason for mistaken assumptions or
rival interpretations. Lakoff and colleagues \cite{Lak00} and Barton \cite{barton} have explored the close connection between language and thought, and shown that images and metaphors used in ordinary language shape mathematical (and all other types of) thinking. We hypothesise that the misconception of a line as a half-line may be due to the naming of the concept; which triggered images of windmills with sails which pivoted around a central tower and extended in one direction only.\footnote{The IMO presents tremendous opportunity for cultural and linguistic analysis, as each problem is translated into at least five different languages, and candidate problems are evaluated partially for the ease with which they can be translated, and the process of translating a problem is taken extremely seriously.}

We expect the use and development of online discussion  to
provide researchers into mathematical practice with large new bodies
of data of informal reasoning in the wild. While it is an open
question whether online mathematics is representative of other
mathematical activity, it is certainly the case that this is one type
of activity. This is validated by peer reviewed collective publications
arising out of online discussions and by the user-base of 23,000 people
on MathOverflow (a small but significant proportion of the world's
research mathematicians).\footnote{Estimates vary from \~{}80,000 (an estimate
  by Jean-Pierre Bourguignon based on the number of people who are in a profession which attaches importance to mathematics research and hold a Mathematics PhD or equivalent \cite{bourguignon}), to \~{}140,000
  (the number of people in the Mathematics Genealogy Project who got their
  PhD between 1960-2012), to \~{}350,000 (the number of people estimated
  still living, on the Math Reviews authors database): see
http://mathoverflow.net/questions/5485/how-many-mathematicians-are-there} It
is also an open question as to whether it is {\em desirable} for
online mathematical collaboration to model offline work, given the
new potential of the online world. As a form of mathematical
practice, it will inform (evolving) theories of (evolving) mathematical practices and -- crucially -- provides a much-needed way of empirically evaluating them. 

The interdisciplinary study of mathematical practice is still very
young, particularly when considered relative to its older, more
respectable sibling, the philosophy of mathematics (\~{}70 years versus
\~{}2,300 years).\footnote{We calculated the 2325 year age gap based on
  Polya's \cite{polya} in 1945 marking the beginning of MP and Plato's
  \cite{plato} in 380 BC on the theory of
  forms and the status of mathematical objects, marking the beginning
  of PoM.} Different disciplines will focus on
different aspects of the sites:  philosophers will concern themselves with
their fundamental question of how mathematics progresses; sociologists on the dynamics of the discussion and the
socio-cultural-technical context in which it takes place; linguists
may analyse the language used, and compare it to other forms of
communication; mathematicians might reflect on whether there is a significant
difference between massively collaborative maths and ordinary
mathematics research; cognitive scientists will look for evidence of
hypothesised cognitive behaviours, and so on. However, these questions
are deeply interrelated. We predict that multi-disciplinary
collaboration in constructing theories of mathematical practice will
increase, and that online discussion sites will play an important role
in uncovering processes and mechanisms behind informal mathematical
collaboration. There is a an exciting potentially symbiotic relationship-in-the-making between
the study of mathematical practice and that of computer support for mathematics.

%

\section{Mathematics as a social machine: the next steps}
The goal of social machines research is to understand the underlying computational and social principles,  and devise a framework for building and deploying them.  

While {\it polymath} and \mo are fairly recent,  the widely used ``Online Encyclopaedia of Integer Sequences''  (www.oeis.org) is a more long-standing example of a social machines for mathematics. Given a few digits of input, it proposes sequences which match it, through invoking subtle pattern matching against over 220,000 user-provided sequences: so, for example, user input of (3 1 4 1) returns $\pi$, and 556 other possibilities, each supported by links to the mathematical literature. Viewed as a social machine, it involves users with queries or proposed new entries; a wiki for discussions; volunteers curating  the system; governance and funding mechanisms through a trust; alongside traditional computer support for a database, matching engine and web interface, with links to other mathematical data sources, such as research papers. While anyone can use the system, proposing a new sequence requires registration and a short CV, which is public, serving as a reputation system.   

One can imagine many kinds of mathematics social machines: the kinds of parameters to be considered in thinking about them in a uniform way include, for example:
\begin{itemize}
\item precise versus loose queries and knowledge
\item human versus machine creativity 
\item specialist or niche users versus general users 
\item logical precision versus cognitive appeal for output
\item formal versus natural language for interaction
\item checking  versus generating conjectures or proofs
\item formal versus informal proof
\item ``evolution" versus ``revolution" for developing new systems
\item governance, funding and longevity
\end{itemize}
Current social and not-so-social machines occupy many different points in this design space.  Each dimension raises broad and enduring challenges, whether in traditional  logic and semantics, human computer interaction, cognitive science, software engineering or information management. 


\subsection{Mathematical elements}
Likely mathematical elements of a mathematics social machine would include the following, all currently major research activities in their own right.

``Traditional'' machine resources available, include software for symbolic and numeric  mathematics such as GAP or Maple,  theorem provers such as Coq or HOL, and bodies of data and proofs arising from such systems. Our work  highlights the importance of including databases of examples, perhaps incorporating user tagging, and also of being able to mine libraries for data and deductions beyond the  immediate facts they record: see in particular the work of Urban \cite{DBLP:conf/cade/UrbanSPV08} on machine learning from such libraries. The emerging field of  mathematical knowledge management \cite{carette} addresses ontologies and tools for sharing and mining such resources, for example providing ``deep'' search or executable papers. Such approaches should in future be able to provide access to  the mathematical literature, especially in the light of ambitious digitisation plans currently being developed by the American Mathematical Society and the Sloan Foundation \cite{NAS}. 
 
\begin{figure}[h!]
\scalebox{0.37}{\includegraphics{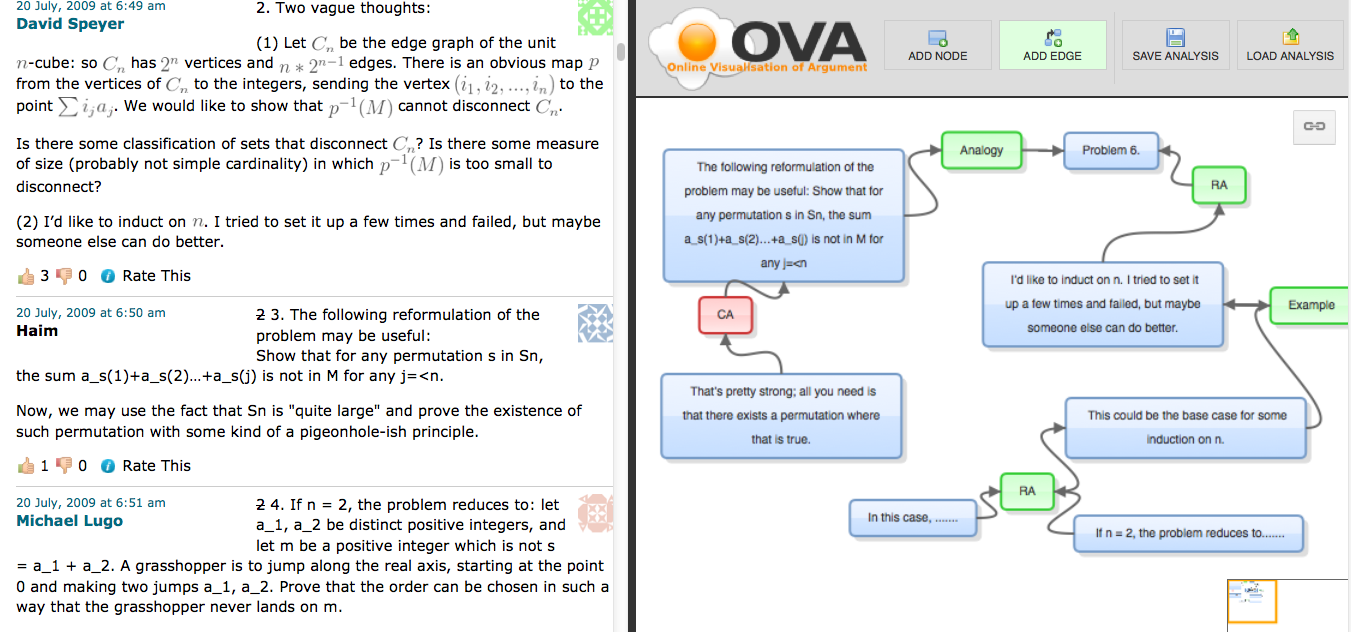}}
\caption{On the left, an extract from the 2009 Mini-polymath conversation, and on the right, a suggested mapping of the discussion using the Online Visualization of
Argument tool developed at the University of Dundee.}
\label{ova}
\end{figure}

The presentation in \mo and {\it polymath} is linear and text based. Machine rendering of mathematical text has
been a huge advance in enabling mathematicians to
efficiently represent their workings {\em in silico}, which in turn
has enabled online
rapid-fire exchange of ideas, but  technology for going beyond the linear structure to capture the more complex structure of a
proof attempt, or to 
represent diagrams, is less developed.  At the end of the
first {\it polymath}  discussion there were 800 comments, and disentangling these for newcomers to the discussion or to write up the proof  for publication can be problematic. Representing the workflow in realtime using argumentation visualization 
software would help prospective participants to more easily understand the
discussion and to more quickly identify areas to which they can
contribute. In Figure \ref{ova} we show two representations of the
mini-Polymath 2009 project\footnote{http://terrytao.wordpress.com/2009/07/20/imo-2009-q6-as-a-mini-polymath-project/}: on the left we see the discussion as it
appeared to participants, and on the right we
have used Online Visualization of
Argument,\footnote{http://ova.computing.dundee.ac.uk/} developed by
Chris Reed and his group at the University of Dundee, to map the argument. 

Turning to the less formal side of mathematics, current challenges raised by the mathematical community, for example see  \cite{Edi2012},  include the importance of collaborative systems that ``think like a mathematician'', can handle unstructured approaches such as the  use of  ``sloppy'' natural language,  support  the exchange of informal knowledge and intuition not recorded in papers, and  engage diverse researchers in creative problem-solving. This mirrors the results of research into mathematical practice:  the importance of human factors, and of handling informal  reasoning,  error, and uncertainty.  Turning messy human knowledge into a usable information space, and reasoning across widely differing user contexts and knowledge bases is only beginning to emerge as a challenge in artificial intelligence applied to mathematics, for example in the work of Bundy  \cite{DBLP:journals/amai/Bundy11} on ``soft'' aspects such as creativity, analogy and concept formation and the handling of error by ontology repair \cite{DBLP:journals/ijswis/McNeillB07}, or work in cognitive science which studies the role of  metaphor in the evolution and understanding of mathematical concepts \cite{Lak00}. 

Automated theory formation systems which automatically invent concepts and conjectures are receiving increasing attention. Examples include
Lenat's AM \cite{lenat:77}, which was designed
to both construct new concepts and conjecture relationships between
them, and Colton's HR system \cite{colton:book,colton:mlj06}. HR uses production rules to form
new concepts from old ones; measures of interestingness
to drive a heuristic search; empirical pattern-based conjecture
making techniques to find relationships between concepts, and 
third party logic systems to prove  conjectures or 
find counterexamples. Other examples include  the
IsaScheme system by Montano Rivas \cite{MontanoRivas2011}, which
employs a scheme-based approach to mathematical theory exploration; the IsaCosy system by Johansson {\em et al.} \cite{johansson}
which performs inductive theory formation by synthesising conjectures
from the available constants and free variables; and the MATHsAiD system by McCasland \cite{mccasland}, which applies inference rules to user-provided axioms, and classifies the resulting proved statements as facts (results of no intrinsic mathematical interest), lemmas
(statements which might be useful in the proof of subsequent
theorems), or theorems (either routine or significant
results). A survey of next generation automated theory
formation is given in \cite{atf-the-next-generation}, including Pease's philosophically-inspired system HRL \cite{pease07}, which provides a 
computational representation of 
Lakatos's theory \cite{lakatos},  and 
Charnley's 
cognitively-inspired  system \cite{charnleyphd} based on  Baar's theory of the Global Workspace
\cite{baarsgwa1}.

Social expectations in {\it mathoverflow}, and generally in research mathematics, are of a culture of open discussion,  and knowledge is freely shared provided it is attributed: for example, it is common practice in mathematics to make papers available before journal submission.  As with mathematics as a whole, information accountability in principle in a mathematics social machine comes from a shared understanding that the arguments presented, while informal, are capable of refinement to a rigorous proof. In {\it mathoverflow}, as described in \cite{DBLP:conf/chi/TausczikP11}, social expectation and information accountability are strengthened through the power of off-line reputation: users are encouraged to use real names, and are likely to interact through professional relationships beyond {\it mathoverflow}.  A further challenge for social computation will be scaling these factors up to larger more disparate communities who have less opportunity for real-world interaction;  dealing in a principled way with credit and attribution as the contributions that social computation systems make  
become routinely  significant; and incorporating models where contributions are traded rather than freely given.

\subsection{Social machines: the broader context}
The research agenda laid out by social machines pioneers like Hendler, Berners Lee and Shadbolt is ambitious \cite{DBLP:journals/ai/HendlerB10}, with a goal of devising overarching principles to understand, design, build and deploy social machines. Viewing mathematics social machines in this way has the potential to  provide a unifying framework for disparate ideas and activities.

 {\it Designing social computations.} Social machine models view users as ``entities" (cf agents or peers) and allow consideration of social interaction, enactment across the network, engagement and incentivisation, and methods of software composition that take into account evolving  social  aggregation.  For mathematics this has far reaching implications ---  handling known patterns of practice, and enabling others as yet unimagined, as well as handling issues such as error and uncertainty, and variations in user beliefs.

{\it Accessing data and information.} Semantic web technology enables databases supporting provenance, annotation, citation and sophisticated search. Mathematics data includes papers,  records of mathematical objects from systems such as Maple, and scripts from theorem provers. There has been considerable research in mathematical knowledge management \cite{DBLP:journals/mics/KohlhaseR12},  but  current experiments in social machines for mathematics  have little such support.  Yet effective search, mining and data re-use would transform both mathematics research and related areas of software verification. Research questions are both technical, for example tracking provenance or ensuring annotation remains timely and correct \cite{DBLP:journals/mscs/CheneyAA11},  and social, for example many {\it mathoverflow} responses cite published work, raising the question of  why users prefer asking {\it mathoverflow} to using a search engine.

 {\it Accountability, provenance and  trust.} Participants in social machines need to be able to trust the processes and data they engage with and share. Key concepts are {\it provenance}, knowing how data and results have been obtained, which contributes to {\it accountability}, ensuring that  the source of any breakdown in trust can be identified and mitigated \cite{Weitzner:2008:IA:1349026.1349043}.  There is a long  tradition of openness in mathematical research which has made endeavours like {\it polymath} or the {\it arXiv} possible and effective 
 --- for example  posting drafts  on the {\it arXiv} ahead of journal submission is reported as speeding up refereeing and reducing priority disputes \cite{Edi2012}.  Trusting mathematical results  requires considering provenance of the proof, a major issue in assessing the balance between formal and informal proofs, and the basis for research into proof certificates \cite{Miller}. Privacy and trust are significant for commercial or government work, where revealing even broad interests  may already be a security concern. 

 {\it Interactions among people, machines and data.} Interactions among people, machines and data are core to social machines, which have potential to support novel forms of interaction and workflow which go beyond current practice, a focus of current social machine research \cite{DBLP:journals/ai/HendlerB10}. Social mathematics shows a variety of 
communities, interactions and purposes, looking for information, solving problems, clarifying information and so on, displaying much broader interactions than those supported by typical mathematical software.  In particular such workflows need to take account of informality and mistakes \cite{DBLP:journals/ws/LehmannVB12}.



In conclusion, social machines both provide new ways of doing mathematics and the means for evaluating theories of mathematical practices. Improved knowledge of human interactions and reasoning in mathematics will suggest new ways in which artificial intelligence and computational mathematics can intersect with mathematics. We envisage that the challenges raised will include developing better computational support for mathematicians and modelling soft aspects of mathematical thinking such as errors, concept development and value judgements. There is much to be done, and a substantial body of research lies ahead of us, but the outcomes could transform the nature and production of mathematics.

\section*{Acknowledgements}
Both authors acknowledge EPSRC support from EP/H500162 and EP/F02309X, and Pease in addition from EP/J004049.  We both thank the School of Informatics at the University of Edinburgh for kind hospitality, where this work was done while the first author was on sabbatical, and thank, in particular, Alan Bundy and members of the DReaM group, and Robin Williams and members of the Social Informatics group, for many helpful discussions and insights.


\end{document}